\documentclass[a4paper,11pt]{article}
\pdfoutput=1
\usepackage{jheppub}
\usepackage[T1]{fontenc}
\usepackage{graphicx}  
\usepackage{dcolumn}   
\usepackage{bm}        
\usepackage{amssymb}   
\usepackage{amsmath}
\usepackage{mathrsfs}
\usepackage{color}
\usepackage[textsize=scriptsize]{todonotes}
\usepackage{lineno}
\usepackage{overpic}
\usepackage{hyperref}
\usepackage{ulem}

\newcommand{\mev}{\mathrm{MeV}}

\newcommand{\mevcc}{\mathrm{MeV}/c^2}
\newcommand{\gev}{\mathrm{GeV}}

\newcommand{\gevcc}{\mathrm{GeV}/c^2}
\newcommand{\jtoss}{J/\psi \rightarrow \Sigma^{+}\overline{\Sigma}^{-}}
\newcommand{\ptoss}{\psi(3686) \rightarrow \Sigma^{+}\overline{\Sigma}^{-}}

\title{\boldmath Measurement of Branching Fractions of $J/\psi$ and $\psi(3686)$ decays to $\Sigma^{+}$ and $\overline{\Sigma}^-$}
\author{
\begin{small}
M.~Ablikim$^{1}$, M.~N.~Achasov$^{10,b}$, P.~Adlarson$^{67}$, S. ~Ahmed$^{15}$, M.~Albrecht$^{4}$, R.~Aliberti$^{28}$, A.~Amoroso$^{66A,66C}$, M.~R.~An$^{32}$, Q.~An$^{63,49}$, X.~H.~Bai$^{57}$
% ...
\\
\begin{center}
\vspace{0.2cm}
(BESIII Collaboration)\\
\end{center}
\vspace{0.2cm} {\it
$^{1}$ Institute of High Energy Physics, Beijing 100049, People's Republic of China\\
% ...
$^{4}$ Bochum Ruhr-University, D-44780 Bochum, Germany\\
% ...
$^{10}$ G.I. Budker Institute of Nuclear Physics SB RAS (BINP), Novosibirsk 630090, Russia\\
% ...
$^{15}$ Helmholtz Institute Mainz, Staudinger Weg 18, D-55099 Mainz, Germany\\
% ...
$^{28}$ Johannes Gutenberg University of Mainz, Johann-Joachim-Becher-Weg 45, D-55099 Mainz, Germany\\
% ...
$^{32}$ Liaoning Normal University, Dalian 116029, People's Republic of China\\
% ...
$^{49}$ State Key Laboratory of Particle Detection and Electronics, Beijing 100049, Hefei 230026, People's Republic of China\\
% ...
$^{57}$ University of Jinan, Jinan 250022, People's Republic of China\\
% ...
$^{63}$ University of Science and Technology of China, Hefei 230026, People's Republic of China\\
% ...
$^{66}$ University of Turin and INFN, (A)University of Turin, I-10125, Turin, Italy; (B)University of Eastern Piedmont, I-15121, Alessandria, Italy; (C)INFN, I-10125, Turin, Italy\\
$^{67}$ Uppsala University, Box 516, SE-75120 Uppsala, Sweden\\
% ...
$^{b}$ Also at the Novosibirsk State University, Novosibirsk, 630090, Russia\\
% ...
}
\vspace{0.4cm}
\end{small}
}

\abstract{Using $1310.6\times10^{6}$ $J/\psi$ and $448.1\times10^{6}$
  $\psi(3686)$ events collected with the BESIII detector, the
  branching fractions of $J/\psi$ decays to
  $\Sigma^{+}\overline{\Sigma}^{-}$ is measured to be $(10.61 \pm
  0.04 \pm 0.36) \times 10^{-4}$, which is significantly more precise than the current world average.
  The branching fractions of $\psi(3686)$ decays to
  $\Sigma^{+}\overline{\Sigma}^{-}$ is measured to be
  $(2.52 \pm 0.04 \pm 0.09) \times
  10^{-4}$, which is consistent with the previous measurements. In addition, the ratio of
  $\mathcal{B}(\psi(3686) \rightarrow
  \Sigma^{+}\overline{\Sigma}^{-})/\mathcal{B}(J/\psi \rightarrow
  \Sigma^{+}\overline{\Sigma}^{-})$ is determined to be $(23.8 \pm
  1.1)\%$ which violates the ``$12\%$ rule''.}

\begin{document} 
\maketitle
\flushbottom

\section{Introduction}
Measurements on the decays of $J/\psi$ and $\psi(3686)$ (denoted here collectively as $\Psi$) can be used to study flavor-SU(3) symmetry breaking and
test quantum chromodynamics (QCD) in the perturbative energy
regime~\cite{pQCD}.  If we consider $J/\psi$ decays into
$B_8\overline{B}_8$ and $B_{10}\overline{B}_{10}$ final states, where
$B_8$ and $B_{10}$ represent the baryon octet and decuplet states,
respectively, and if the electromagnetic contributions are neglected,
flavor-SU(3) symmetry gives the same decay amplitudes for all $J/\psi$
decays to baryon anti-baryon pairs.  However, broken flavor-SU(3)
symmetry can contribute to the differences in branching fractions of
different baryonic pairs.  Furthermore, the branching fractions are
determined not only by strong interaction amplitudes, but also by
electromagnetic interactions and interferences between
them~\cite{Rudaz:1975yu}, although these are much smaller than the expected flavor-SU(3) breaking effects. As shown in Table~\ref{tab:theory}, a phenomenologically plausible model~\cite{Zhu:2015bha,Ferroli:2019nex} can be made to fit the pattern of branching fractions of $J/\psi$ decays to baryon octet final states well~\cite{pdg2018}. However
the precision on the branching fraction of $J/\psi \rightarrow
\Sigma^{+}\overline {\Sigma}^{-}$ is still relatively
poor~\cite{Ablikim:2008tj}.  The $1310.6 \times 10^{6}$ $J/\psi$ event
sample collected by the BESIII experiment in 2009 and 2012 allows a
much more precise measurement of the branching fraction of $J/\psi
\rightarrow \Sigma^{+}\overline{\Sigma}^{-}$.

\begin{table}[h]
\vspace{-2mm}
\centering
\caption{Comparison of the experimental measurements
  ($\mathcal{B}_{\rm pdg}$)~\cite{pdg2018} and phenomenological
  calculations ($\mathcal{B}_{\rm
    cal}$)~\cite{Zhu:2015bha,Ferroli:2019nex} for the branching
  fractions of $J/\psi$ decays to baryon octet final states, where
  $\Delta(\sigma)$ is the difference in terms of the total
  uncertainty. Dash ($-$) represents no experimental measurement.}
\label{tab:theory} 
\begin{tabular}{lrrr} 
\hline\noalign{\smallskip}
$B\overline{B}$ & $\mathcal{B}_{\rm pdg} (10^{-3})$ & $\mathcal{B}_{\rm cal} (10^{-3})$ & $\Delta(\sigma)$ \\
\noalign{\smallskip}\hline\noalign{\smallskip}%
$\Sigma^0 \overline \Sigma^0$ & $1.164 \pm 0.004 $ & $1.160 \pm 0.041 $ & $\sim 0.09$ \\
$\Lambda \overline \Lambda$ & $1.943 \pm 0.003 $ & $1.940 \pm 0.055 $ & $\sim 0.05$ \\
$\Lambda \overline \Sigma^0 +$ c.c. & $0.0283 \pm 0.0023$ & $0.0280 \pm 0.0024$ & $\sim 0.06$ \\
$p \overline p$ & $2.121 \pm 0.029 $ & $2.10 \pm 0.16 $ & $\sim 0.1$ \\
$n \overline n$ & $2.09 \pm 0.16 $ & $2.10 \pm 0.12 $ & $\sim 0.04$ \\
$\Sigma^+ \overline \Sigma^-$ & $1.50 \pm 0.24 $ & $1.110 \pm 0.086 $ & $\sim 1$ \\
$\Sigma^- \overline \Sigma^+$ & $-$ & $0.857 \pm 0.051 $ & $-$ \\
$\Xi^0 \overline \Xi^0$ & $1.17 \pm 0.04 $ & $1.180 \pm 0.072 $ & $\sim 0.09$ \\
$\Xi^- \overline \Xi^+$ & $0.97 \pm 0.08 $ & $0.979 \pm 0.065 $ & $\sim 0.06$ \\
\noalign{\smallskip}\hline
\end{tabular}
\end{table}%

According to pQCD, the ratio of $\Gamma_{h}$ to $\Gamma_{l}$, where
$\Gamma_{h}$ is the partial width of $J/\psi$ ($\psi(3686)$) decay to
light hadrons and $\Gamma_{l}$ is the partial width to leptons, does
not depend on the particle wave function~\cite{12rule}.  The ratio
between the branching fractions of $J/\psi$ and $\psi(3686)$ decays to
the same final states obeys the so-called ``$12\%$
rule'',
\begin{equation*}
\frac{\mathcal{B}_{\psi(3686)\to h}}{\mathcal{B}_{J/\psi\to h}} \approx \frac{\mathcal{B}_{\psi(3686)\to l^+l^-}}{\mathcal{B}_{J/\psi\to l^+l^-}} = (13.3 \pm 0.3)\%.
\end{equation*}
Although a large fraction of exclusive decay channels follow the rule
approximately, significant violation was first observed in the $\rho
\pi$ channel~\cite{Franklin:1983ve}. The ratio of $\mathcal{B}(\psi(3686) \to \rho \pi)$ to
$\mathcal{B}(J/\psi \to \rho \pi)$ is much smaller than the pQCD prediction, and this is called the
``$\rho\pi$'' puzzle. 
To understand the ``$\rho\pi$'' puzzle, the theoretical and experimental efforts have been made: amongst the suggested solutions are $J/\psi$-glueball admixture scheme, Instrinsic-charm-component scheme, Squential-fragmentation model, Exponential-form-factor model, S-D wave mixing scheme, Final state interaction scheme and so on~\cite{Ref:review}.
But there are no satisfactory explanations for all existing experimental results. 
Tests of the $12\%$ rule using the baryonic
decay modes may be helpful in understanding the $\rho \pi$ puzzle.
With CLEO data~\cite{Pedlar:2005px, Dobbs:2017hyd}, the
branching fraction of $\psi(3686) \rightarrow \Sigma^{+}
\overline{\Sigma}^{-}$ was determined to be $(2.32 \pm 0.12) \times
10^{-4}$.  The BESIII experiment, having collected the largest sample
of $\psi(3686)$ events, gives the opportunity to improve the precision
of this branching fraction and test the $12\%$ rule.

In this paper we report, with improved precision, the branching
fraction measurements of $\jtoss$ and $\ptoss$ based on
$1310.6\times10^{6}$ $J/\psi$ and $448.1\times10^{6}$ $\psi(3686)$
events collected with the BESIII detector at the BEPCII collider
during 2009 and 2012.
\section{BESIII Detector and Monte Carlo Simulation}

The BESIII detector~\cite{Ablikim:2009aa} records symmetric $e^+e^-$
collisions provided by the BEPCII storage
ring~\cite{Yu:IPAC2016-TUYA01}, which operates with a peak luminosity
of $1\times10^{33}$~cm$^{-2}$s$^{-1}$ in the center-of-mass energy
range from 2.0 to 4.9~GeV.  BESIII has collected large data samples in
this energy region~\cite{Ablikim:2019hff}. The cylindrical core of the
BESIII detector covers 93\% of the full solid angle and consists of a
helium-based multilayer drift chamber~(MDC), a plastic scintillator
time-of-flight system~(TOF), and a CsI(Tl) electromagnetic
calorimeter~(EMC), which are all enclosed in a superconducting
solenoidal magnet providing a 1.0~T (0.9~T in 2012) magnetic
field. The solenoid is supported by an octagonal flux-return yoke with
resistive plate counter muon identification modules interleaved with
steel.  The charged-particle momentum resolution at $1~{\rm GeV}/c$ is
$0.5\%$, and the $dE/dx$ resolution is $6\%$ for electrons from Bhabha
scattering. The EMC measures photon energies with a resolution of
$2.5\%$ ($5\%$) at $1$~GeV in the barrel (end cap) region. The time
resolution in the TOF barrel region is 68~ps, while that in the end
cap region is 110~ps.

Monte Carlo~(MC) simulated events are used to determine the detection
efficiency, optimize selection criteria, and study possible
backgrounds.  GEANT4-based~\cite{geant4} MC simulation software,
which includes the geometric and material descriptions of the BESIII
detector, the detector response, and digitization models as well as
the detector running conditions and performance, is used to generate
MC samples. The simulation models the beam
energy spread and initial state radiation (ISR) in the $e^+e^-$
annihilations with the generator {\sc
kkmc}~\cite{KKMC}. The inclusive MC samples of $J/\psi$ and $\psi(3686)$ includes the production of the
$J/\psi$ and $\psi(3686)$ resonances, the ISR production of the $J/\psi$, and
the continuum processes incorporated in {\sc
kkmc}. The known decay modes are modelled with {\sc
evtgen}~\cite{evtgen} using branching fractions taken from the
Particle Data Group~\cite{pdg}, and the remaining unknown charmonium decays
are modelled with {\sc lundcharm}~\cite{lundcharm}. Final state radiation~(FSR)
from charged final state particles is incorporated using the {\sc
photos} package~\cite{photos}. To describe
the MC simulation of the signal process, the differential cross
section is expressed with respect to five observables
$\boldsymbol{\xi}= ( \theta_{\Sigma^{+}}, \theta_{p}, \phi_{p},
\theta_{\overline{p}},
\phi_{\overline{p}})$~\cite{Faldt:2017kgy}. Here $\theta_{\Sigma^{+}}$
is the angle between the $\Sigma^{+}$ and electron ($e^{-}$) beam in
the interaction center-of-mass frame (CM), $\theta_{p}, \phi_{p}$ and
$\theta_{\overline{p}}, \phi_{\overline{p}}$ are the polar and
azimuthal angles of the proton and anti-proton measured in the rest
frames of their corresponding mother particles.  The parameters in the
differential cross sections have been determined in
Ref.~\cite{Ablikim:2020lpe}.

\section{Selection criteria}

Candidates of $\Psi \rightarrow \Sigma^{+} \overline{\Sigma}^{-}$,
where $\Sigma^{+}$($\overline{\Sigma}^{-})\rightarrow
p\pi^{0}$($\overline{p}\pi^{0}$) and $\pi^{0} \rightarrow \gamma
\gamma$ are required to have two charged tracks with opposite
charges and at least four photons. Charged tracks are required to be
within the acceptance of the MDC.  For each track, the point of
closest approach to the interaction point must be within 2~cm in the
plane perpendicular to the $z$ axis and within $\pm$10~cm along $z$,
where $z$ is along the symmetry axis of the MDC. A particle
identification algorithm~(PID) combines measurements of the energy
deposited in the MDC~(d$E$/d$x$) and the flight time to the TOF to
form likelihoods $\mathcal{L}(h)~(h=p,K,\pi)$ for each hadron
hypothesis. The two good charged tracks are identified as proton and
anti-proton by requiring $\mathcal{L}(p) > \mathcal{L}(\pi)$ and
$\mathcal{L}(p) > \mathcal{L}(K)$.

Photon candidates are reconstructed from isolated showers in the
EMC. Each photon candidate is required to have a minimum energy of
25~$\mev$ in the EMC barrel region or 50~$\mev$ in the end cap
region. To improve the reconstruction efficiency and the energy
resolution, the energy deposited in the nearby TOF counters is
included in the photon reconstruction. To suppress electronic noise
and showers unrelated to the event, the difference between the EMC
time and the event start time is required to be within [0, 700]\,ns.
The $\pi^{0}$ candidates are reconstructed by requiring the invariant
mass of photon pairs to satisfy $(M_{\pi^0}-60) <M_{\gamma \gamma} <
(M_{\pi^0}+40)~\mevcc$, where $M_{\pi^0}$ is the nominal mass of
$\pi^0$~\cite{pdg}. The asymmetrical mass window is used because the
photon energy deposited in the EMC has a tail on the low energy side.
A one-constraint (1C) kinematic fit is performed on the photon pairs
by constraining their invariant masses to the nominal $\pi^0$ mass,
and the $\chi^2_{\rm 1C}$ is required to be less than 25 to remove
fake candidates. Further there must be at least two reconstructed
$\pi^0$ candidates.

To further remove potential background events and improve the mass
resolution, a four-constraint (4C) kinematic fit is performed,
constraining the total reconstructed four momentum to that of the
initial $e^+e^-$ state. A requirement on the quality of the 4C
kinematic fit of $\chi^2_{\rm 4C} < 100$ is imposed, which is chosen
by optimizing the figure-of-merit, defined as $\frac{S}{\sqrt{S+B}}$,
where $S$ is the number of signal events and $B$ is the number of
background events, which are estimated based on MC simulations.  If the number of $\pi^0$ candidates in an event is
greater than two, the $p\overline{p}\pi^{0}\pi^{0}$ combination with
the lowest $\chi^{2}_{\rm 4C}$ is selected.  After kinematic fitting,
the $\Sigma^{+}$ and $\overline{\Sigma}^{-}$ candidates are
constructed from the proton, anti-proton and neutral-pion candidates,
and the combination that minimizes $ \sqrt{(M_{p\pi^{0}}
  -m_{\Sigma^{+}})^2 + (M_{\overline{p}\pi^{0}} -
  m_{\overline{\Sigma}^{-}})^2}$ is chosen in order to match the
neutral pions to the corresponding baryons.  For the $\psi(3686)
\rightarrow \Sigma^{+}\overline{\Sigma}^{-}$ decay, an additional
invariant mass requirement is imposed on the proton-antiproton pair,
$|M_{p\overline{p}} - 3.1| > 0.05$ $\gevcc$, to remove the background
of $\psi(3686) \to \pi^0 \pi^0 J/\psi$ with $J/\psi \rightarrow p
\overline{p}$.

\begin{figure}[b]
\begin{center}
\begin{overpic}[width=0.45\textwidth,angle=0]{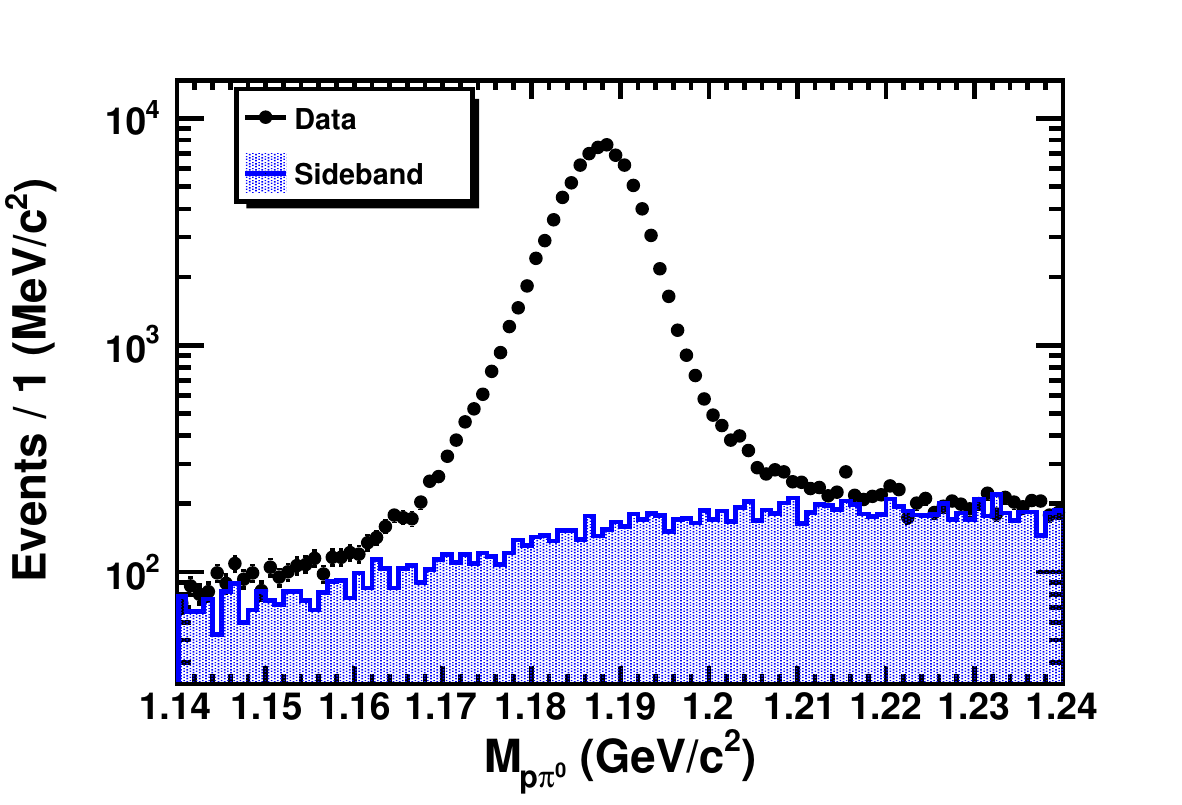}
\put(70, 50){(a)}
\end{overpic}
\begin{overpic}[width=0.45\textwidth,angle=0]{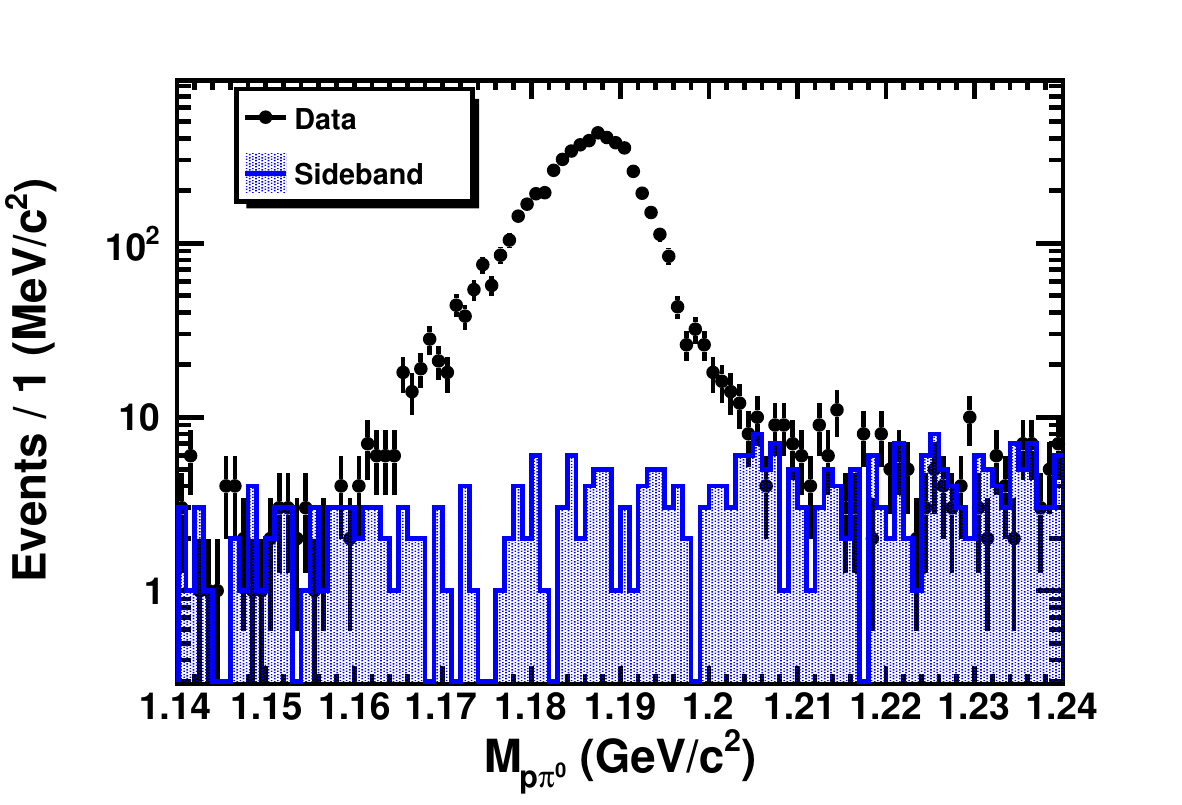}
\put(70, 50){(b)}
\end{overpic}
\begin{overpic}[width=0.45\textwidth,angle=0]{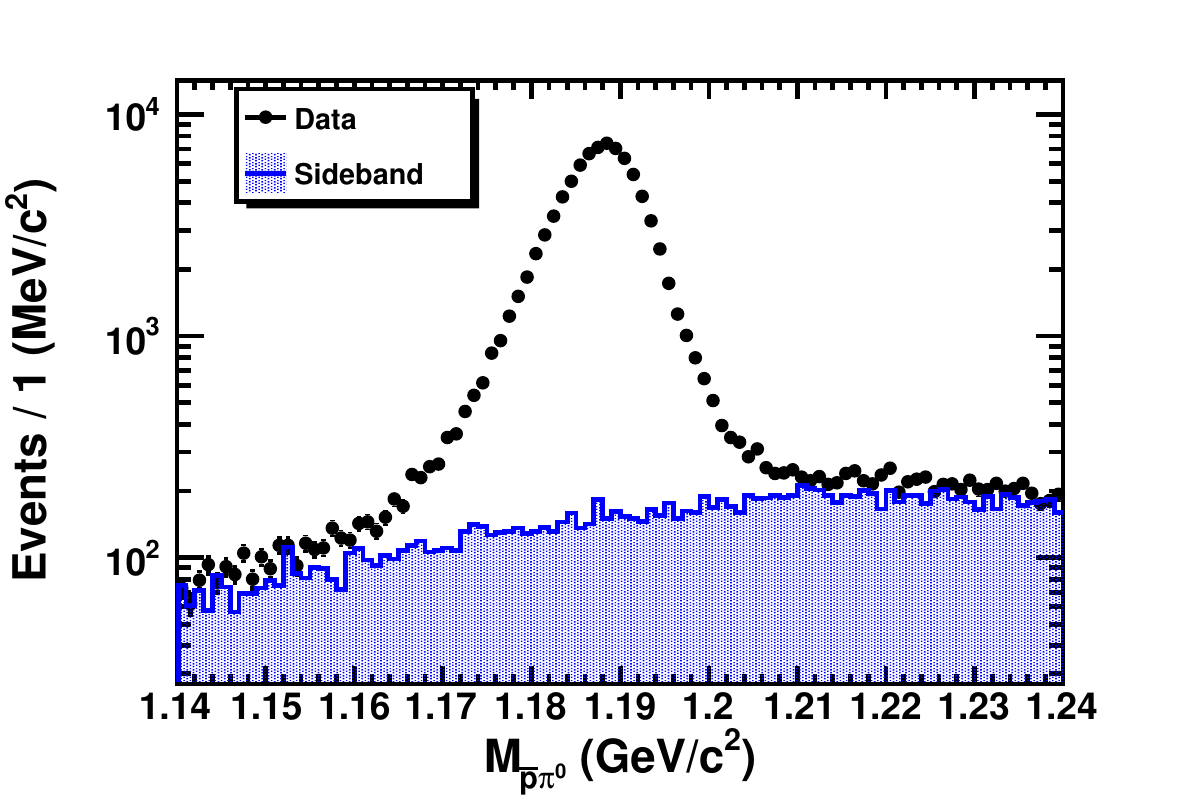}
\put(70, 50){(c)}
\end{overpic}
\begin{overpic}[width=0.45\textwidth,angle=0]{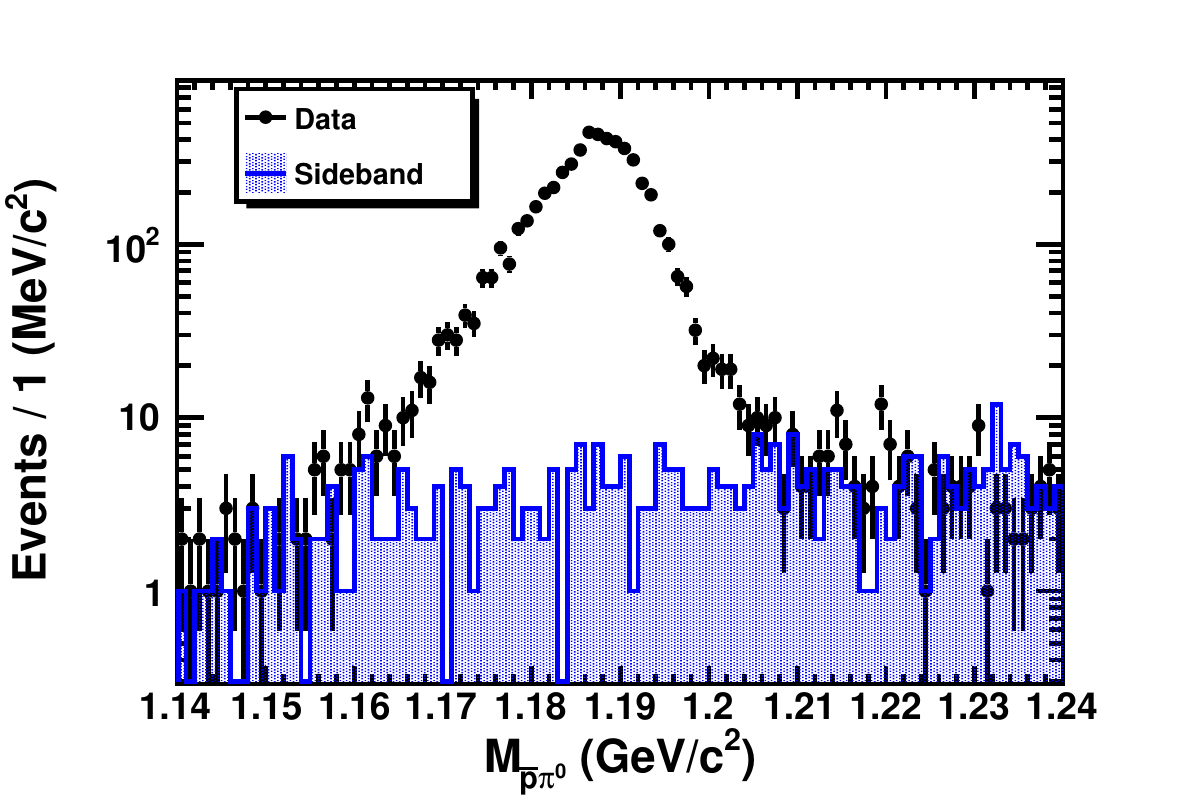}
\put(70, 50){(d)}
\end{overpic}
\end{center}
\caption{The invariant mass distributions of $p \pi^{0}$ (top) by requiring
  $M_{\overline{p}\pi^{0}}$ to be within the signal or sideband
  region and
  $\overline{p} \pi^{0}$ (bottom)by requiring
  $M_{p\pi^{0}}$ to be within the signal or sideband
  region. Left are $\jtoss$ distributions and right are
  $\ptoss$, where the black dots are data and the blue histograms are
  sideband contributions.}
\label{fig:sideband}
\end{figure}

To investigate other possible background processes, inclusive MC
samples of $1.2 \times 10^9$~$J/\psi$ and $5.06 \times
10^8$~$\psi(3686)$ decays are used and
examined TopoAna, a software tool to categorise backgrounds and identify the physics processes of interests
from the inclusive MC samples~\cite{Zhou:2020ksj}.  For $\jtoss$, the
  dominant background contributions are found to be $J/\psi
  \rightarrow \Delta^{+} \overline{\Delta}^{-}$, $J/\psi \rightarrow p
  \overline{p} \pi^{0} \pi^{0}$, $J/\psi \rightarrow \gamma \Sigma^{+}
  \overline{\Sigma}^{-}$ and $J/\psi \rightarrow \gamma \eta_{c}$ with
  subsequent decay $\eta_{c} \rightarrow \Sigma^{+}
  \overline{\Sigma}^{-}$.  For the $\ptoss$, the main background
  contributions are from $\psi(3686) \rightarrow \Delta^{+}
  \overline{\Delta}^{-}$, $\psi(3686) \rightarrow p \overline{p}
  \pi^{0} \pi^{0}$, $\psi(3686) \rightarrow \gamma \chi_{c2}$ with
  $\chi_{c2} \rightarrow p \overline{p} \pi^{0}$ and $\psi(3686)
  \rightarrow \gamma \eta_{c}$ with $\eta_{c} \rightarrow \Sigma^{+}
  \overline{\Sigma}^{-}$.  All the above backgrounds can be classified
  as peaking backgrounds or non-peaking backgrounds depending on
  whether there is $\Sigma^{+}\overline{\Sigma}^{-}$ in the final
  states.  The peaking background contributions are estimated to be
  less than 0.1\% of the signal and can thus be neglected. For
  non-peaking backgrounds, we treat them as a smooth distribution in
  the invariant mass spectrum. The sideband method is used to further
  check the background distributions. The signal region is defined as
  $1.17 < M_{p\pi^{0}/\overline{p}\pi^{0}} < 1.2~\gevcc$, and the
  lower and upper sideband
  regions are defined as $1.14 < M_{p\pi^{0} / \overline{p}\pi^{0}} <
  1.155~\gevcc$ and $1.225 < M_{p\pi^{0} / \overline{p}\pi^{0}} <
  1.24~\gevcc$, respectively.  The $M_{p\pi^{0}}$ distributions, shown
  in Fig~\ref{fig:sideband}(a)(b), are plotted by requiring
  $M_{\overline{p}\pi^{0}}$ to be within the signal or sideband
  region. Similarly, in Fig~\ref{fig:sideband}(c)(d), the
  $M_{\overline{p}\pi^{0}}$ distributions are plotted by requiring
  $M_{p\pi^{0}}$ to be within the signal or sideband region. There are
  no obvious peaking background contributions in the sideband region.

To investigate the contributions from the quantum electrodynamics
(QED) process of $e^{+} e^{-} \rightarrow \Sigma^{+}
\overline{\Sigma}^{-}$, two continuum datasets collected at
center-of-mass energies of 3.08 $\gev$ and 3.65 $\gev$, with
luminosities of 30 $\rm {pb}^{-1}$ and 44 $\rm {pb}^{-1}$, are
used. The absolute magnitude is determined according to the formula
$N = N^{survived}_{continuum} \cdot \frac{\mathcal{L}_{\psi(3686)}}{\mathcal{L}_{continuum}} \cdot \frac{3.65^{2}}{3.686^{2}}$, where $N^{survived}_{continuum} = 2$ is the number of events which remained in the off-resonance sample after applying the same selection criteria, $\mathcal{L}_{\psi(3686)} =$ 668.55 $\rm {pb}^{-1}$ and $\mathcal{L}_{continuum} =$ 44 $\rm {pb}^{-1}$. The contribution of QED process is negligible for our measurement.

\section{Branching fraction measurements} 
\begin{figure}[b] \begin{center}
\begin{overpic}[width=0.45\textwidth,angle=0]{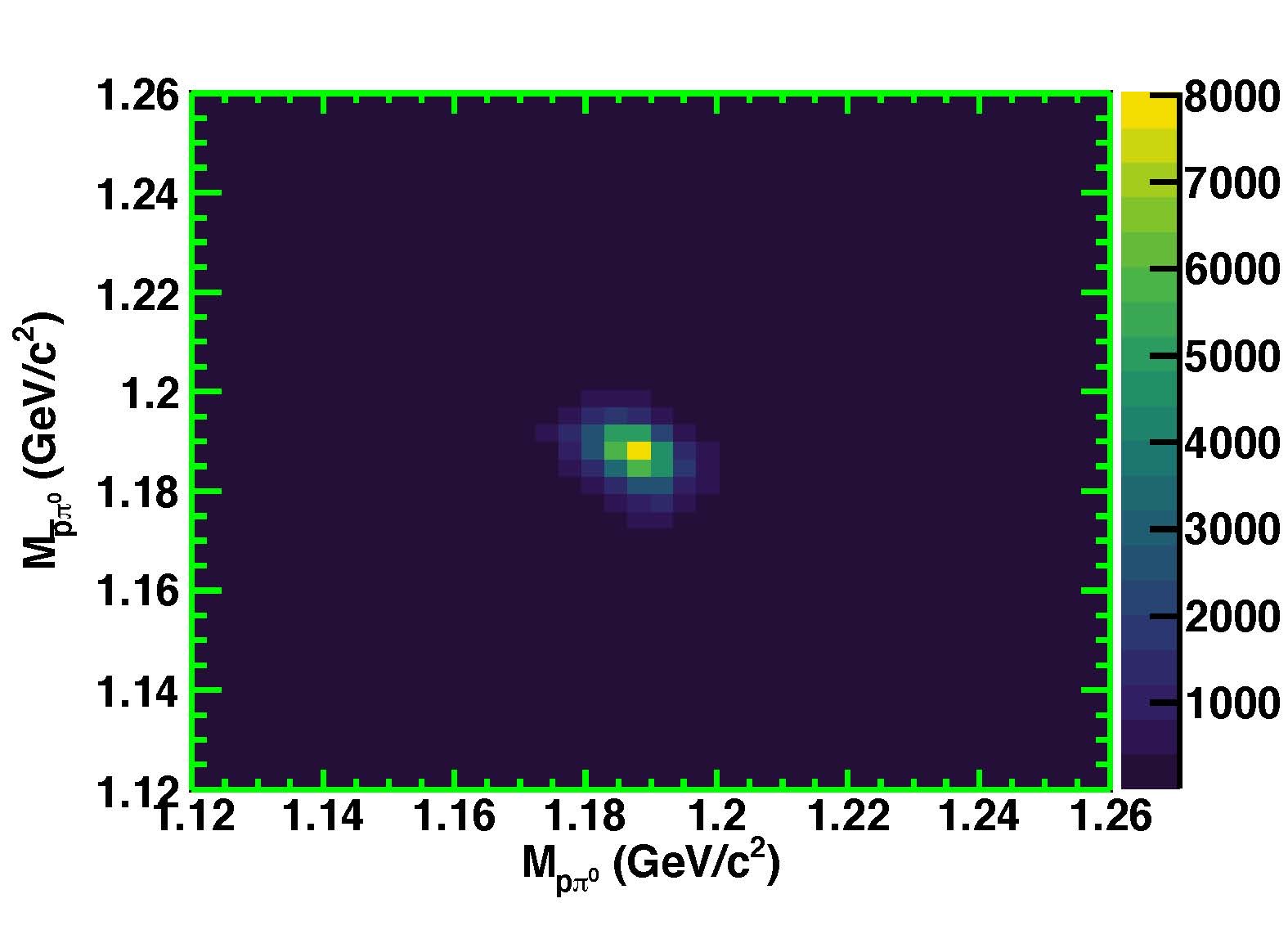}
\end{overpic}
\begin{overpic}[width=0.45\textwidth,angle=0]{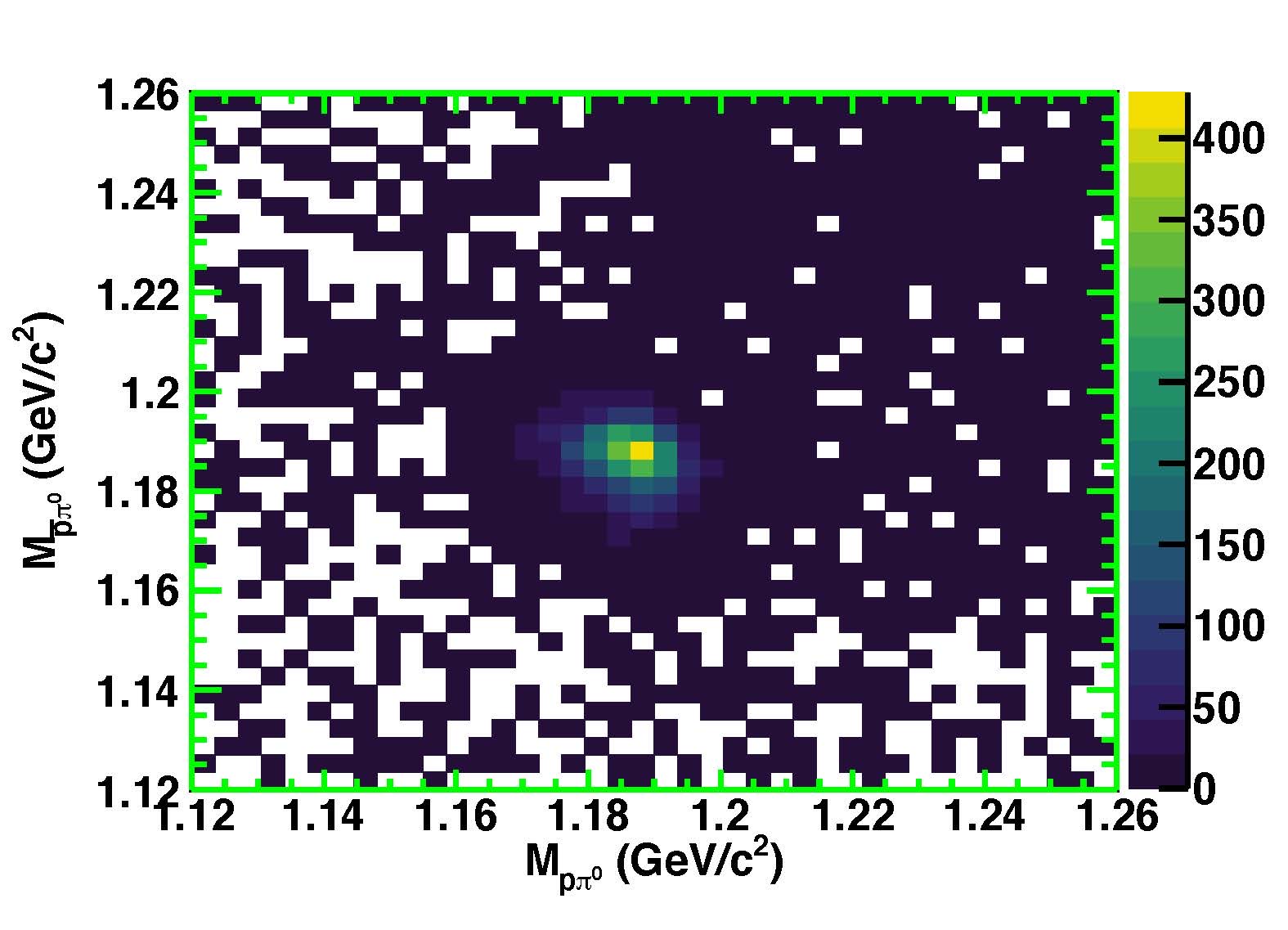}
\end{overpic} \end{center} \caption{The two-dimensional distributions
of $M_{p\pi^{0}}$ versus $M_{\overline{p}\pi^{0}}$. The left is
$J/\psi$ decay, and the right is $\psi(3686)$ decay.}
\label{fig:2dplot} \end{figure}

\begin{figure}[htbp]
\begin{center}
\begin{overpic}[width=0.45\textwidth,angle=0]{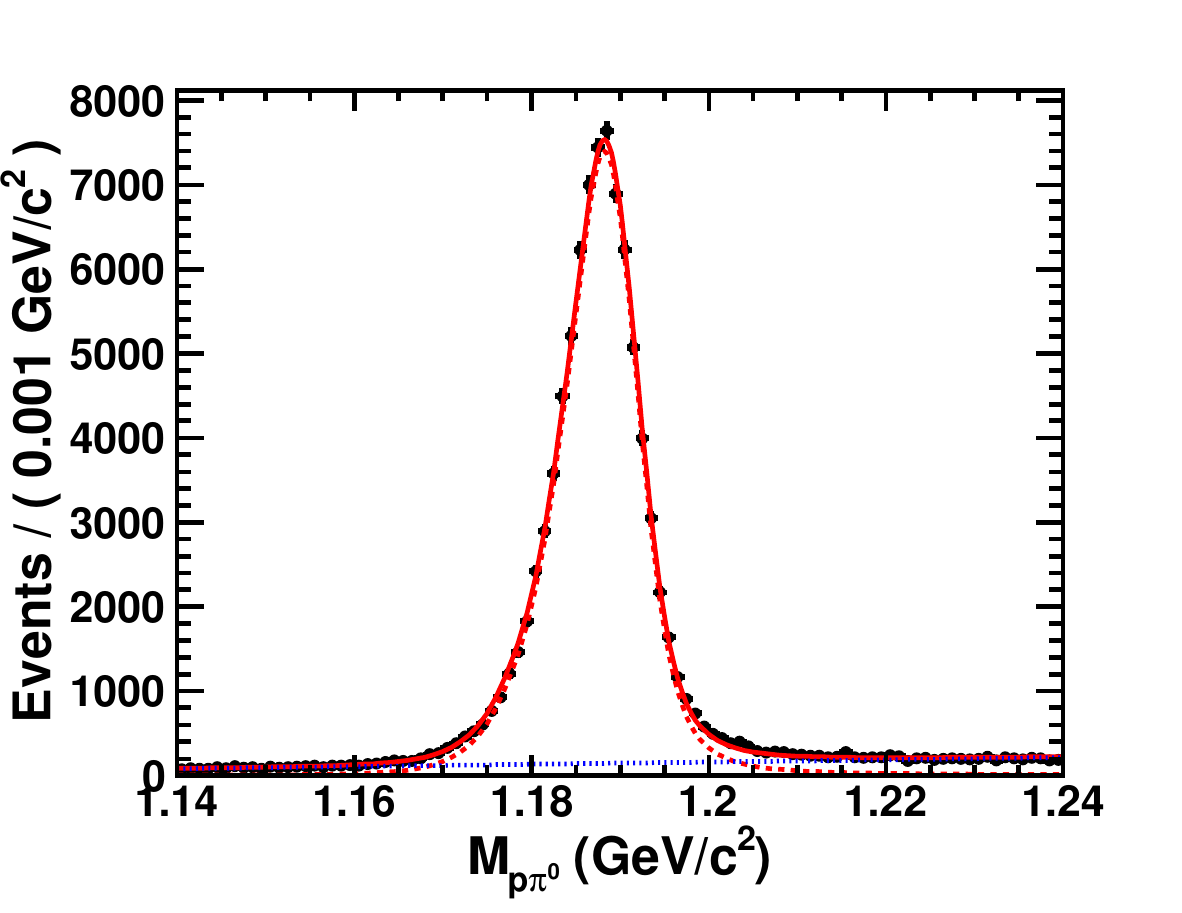}
\end{overpic}
\begin{overpic}[width=0.45\textwidth,angle=0]{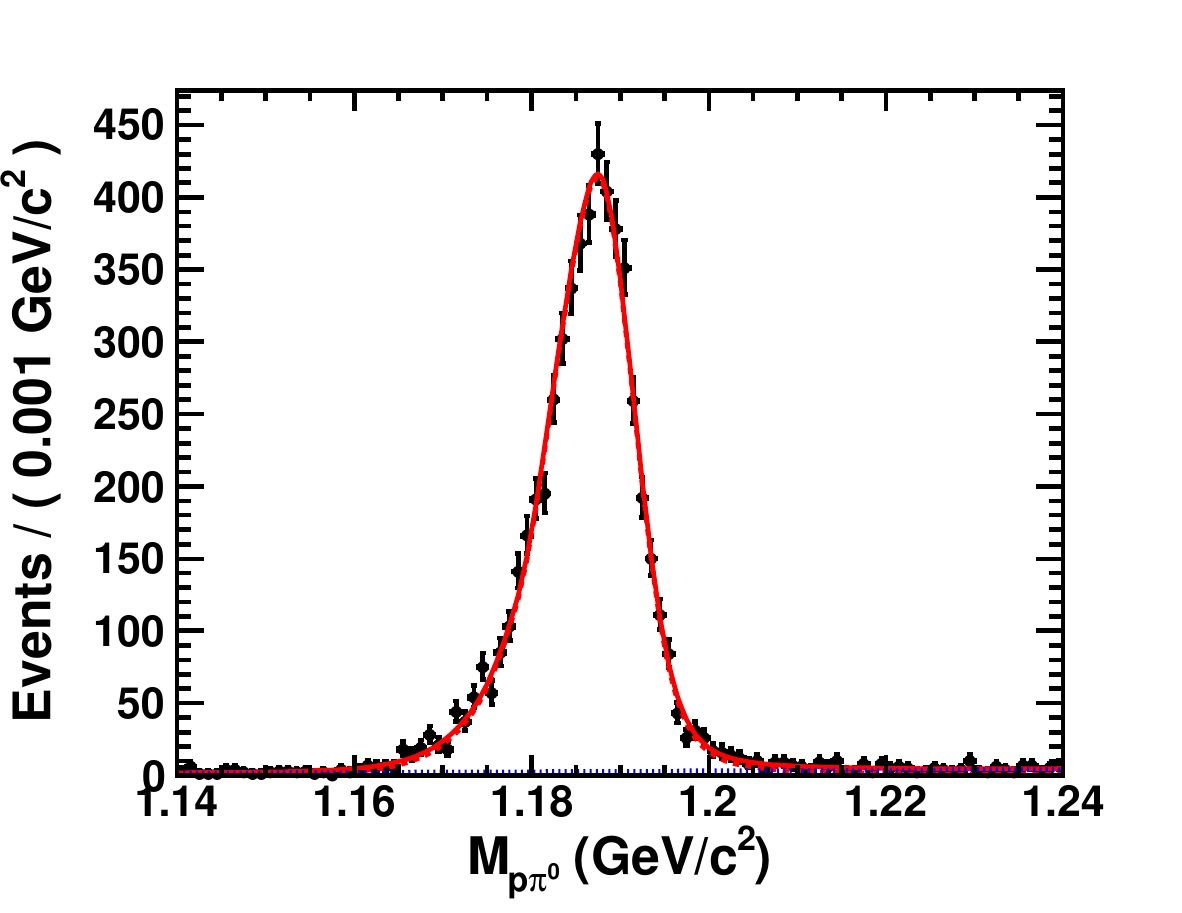}
\end{overpic}
\begin{overpic}[width=0.45\textwidth,angle=0]{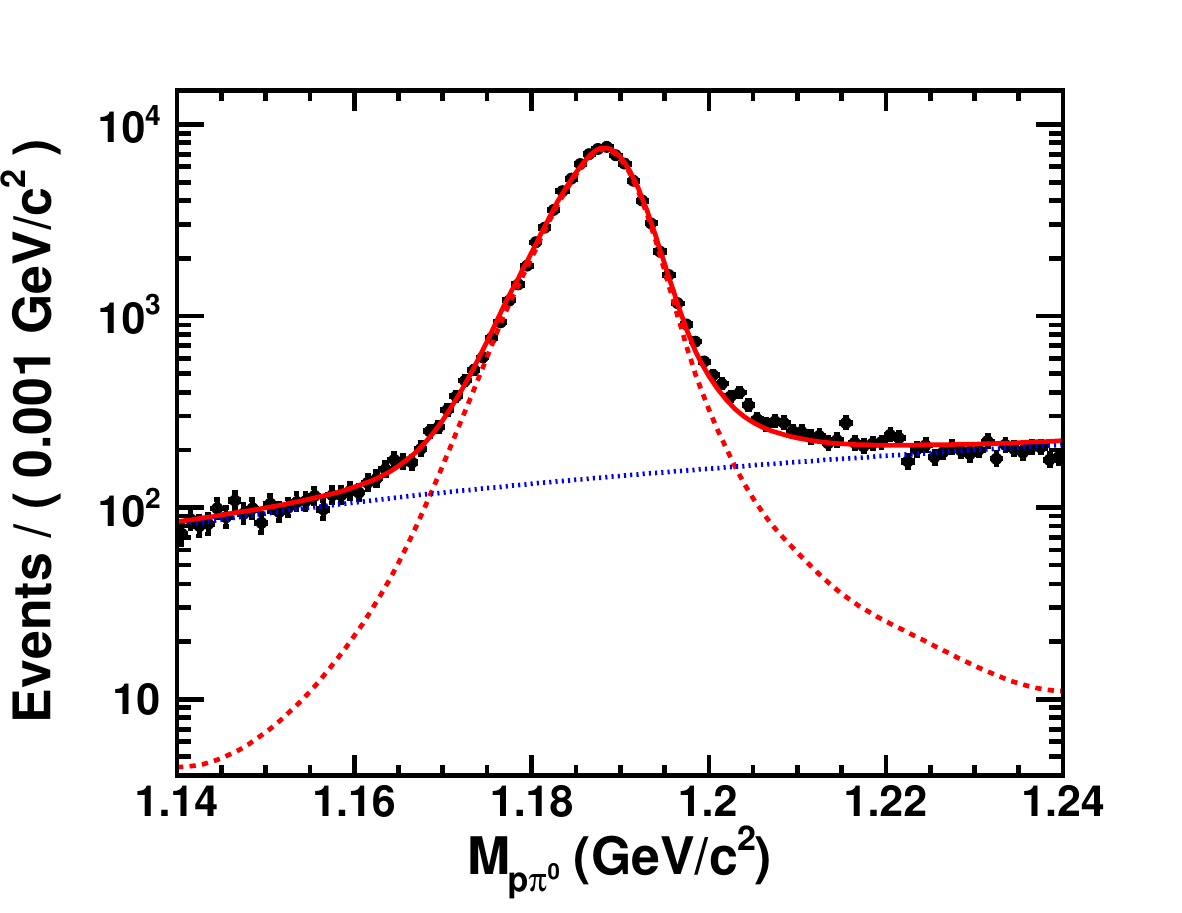}
\end{overpic}
\begin{overpic}[width=0.45\textwidth,angle=0]{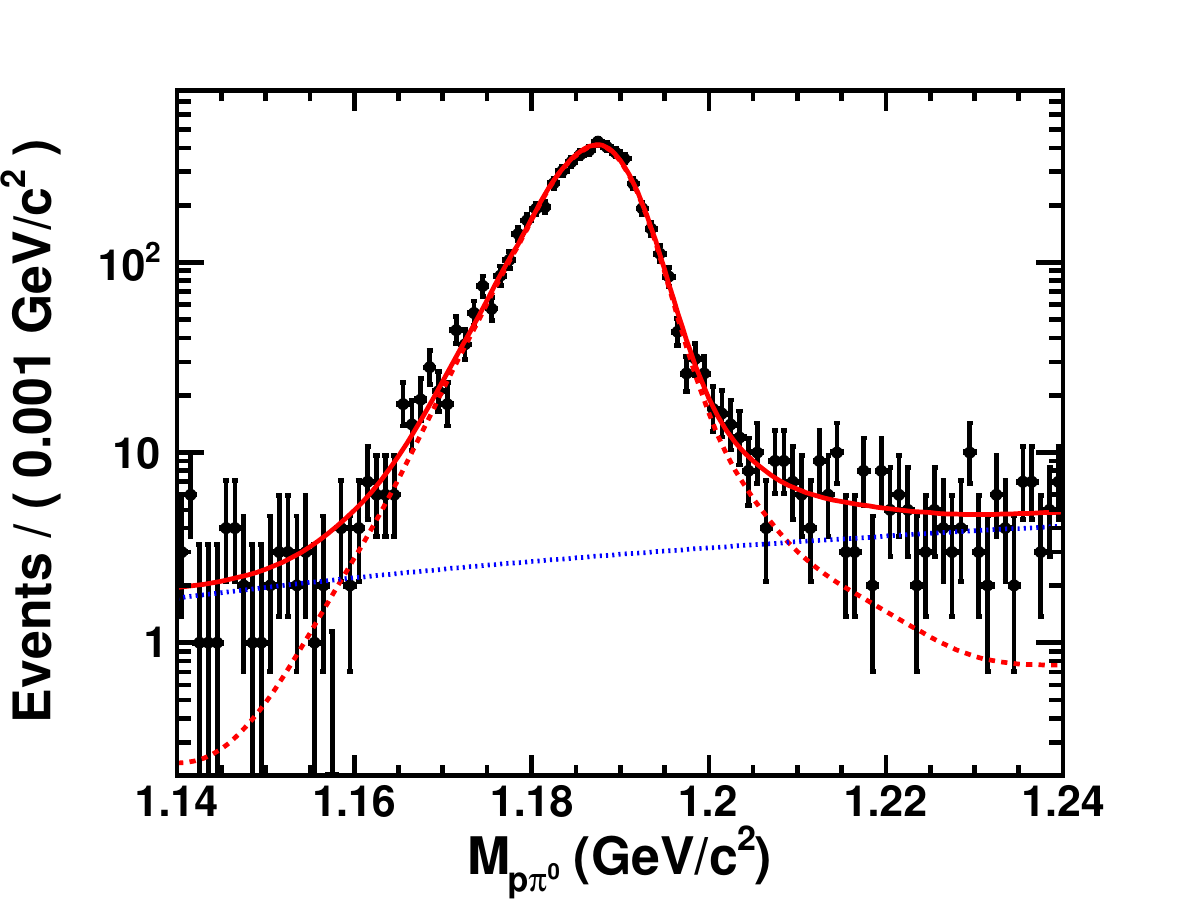}
\end{overpic}
\end{center}
\caption{The $\Sigma^{+}$ invariant mass distributions with fit
  results superimposed. The left are for $\jtoss$ and the right for
  $\ptoss$. The bottom plots are in log scale.  The red solid lines are
  the total fitting functions, the red dashed lines are the signal
  functions and the blue dotted ones are the background functions.  }
\label{fig:fitting_total}
\end{figure}

With the above selection
criteria, there are significant enhancements close to the $\Sigma^{+}$
and $\overline{\Sigma}^{-}$ nominal masses in the two dimensional
distribution of $M_{p\pi^{0}}$ and $M_{\overline{p}\pi^{0}}$ as can be
seen in Fig.~\ref{fig:2dplot}.   To obtain the number of signal events,
an unbinned maximum likelihood fit is performed to the $M_{p\pi^{0}}$ distribution by requiring
$M_{\overline{p}\pi^{0}}$ to be within the signal region.  The signal
is described by the MC shape convoluted with a Gaussian function which
represents the difference between data and MC in the resolution and
mean value.  The background is described with a second-order
polynomial function. The mean and width of the Gaussian function and polynomial function parameters are all floated. Figure~\ref{fig:fitting_total} shows the fitting
of the $p\pi^{0}$ invariant mass distributions, where the red solid
lines are the total fitting functions, the red dashed lines are the
signal functions and the blue dotted ones are the background
functions.

The branching fraction of each channel is calculated according to 
\begin{equation}
\mathcal{B}(\Psi \rightarrow \Sigma^{+} \overline{\Sigma}^{-}) = \frac{N_{sig}}{\epsilon_{\rm cor} \times \Pi\mathcal{B}_{i} \times N_{\Psi}},
\end{equation}
where $N_{sig}$ is the number of signal events determined by the fit,
$\epsilon_{\rm cor}$ is the corrected detection efficiency, generated
according to the decay parameters measured in data but corrected for
differences between data and MC simulation, $\Pi\mathcal{B}_i$ is the
product of the branching fractions of all the intermediate states in
each channel and $N_{\Psi}$ is the number of $J/\psi$ or
$\psi(3686)$ events~\cite{njpsi, npsp}.

The corresponding numbers of signal events, detection efficiencies and
branching fractions are listed in Table~\ref{table:br}.  The initial
detection efficiencies are estimated with signal MC simulation. In the
calculation of $\epsilon_{cor}$, we take into account the difference
between data and signal MC, obtained from control samples, which
include the differences of detection efficiencies of the proton,
anti-proton and $\pi^{0}$.  To study the tracking and PID efficiencies
of the proton and anti-proton, the decay processes of $\Psi
\rightarrow p \overline{p} \pi^{+} \pi^{-}$ are used to select the
control samples of the proton or anti-proton.  The proton efficiency
ratios between MC and data are determined within different proton
transverse-momentum and polar-angle regions.  The ratios of
anti-proton efficiency are also determined using the same method.  To
study the $\pi^{0}$ reconstruction efficiency, the control samples are
selected with the processes of $\psi(3686) \rightarrow \pi^{0} \pi^{0}
J/\psi, J/\psi \rightarrow l^{+} l^{-}$ and $e^{+} e^{-} \rightarrow \omega \pi^{0}$ at $\sqrt{s} = 3.773$ GeV.  The relative
difference of the $\pi^{0}$ reconstruction efficiencies between MC and
data obtained on the two datasets are consistent with each other and
depend on $\pi^0$ momentum.  The overall correction to the event
selection efficiency is the product of correction factors of proton,
anti-proton and $\pi^{0}$ in the related kinematic regions.

For the consistence study of 2009 and 2012 datasets, we study the branching fractions separately (only with statistical uncertainties). For $J/\psi$ decay, the branching fractions are $(10.73 \pm 0.09)\times 10^{-4}$ and $(10.59 \pm 0.04)\times 10^{-4}$ for 2009 and 2012 separately, which are consistent with each other within 1.5 standard deviations. For $\psi(3686)$ decay, the branching fractions are $(2.57 \pm 0.07)\times 10^{-4}$ and $(2.50 \pm 0.04)\times 10^{-4}$ for 2009 and 2012 separately, which are consistent with each other within 1 standard deviations.

\begin{table*}[htbp]
\caption{The numbers of signal events, detection efficiencies and
  branching fractions of $\Psi \rightarrow \Sigma^{+}
  \overline{\Sigma}^{-}$, where the uncertainties are statistical
  only.}
\begin{center}
\begin{tabular}{c|c|c|c}
\hline
\hline
Channel  &$N_{sig}$  &$\epsilon_{cor} (\%)$ &Branching fraction $(10^{-4})$ \\
\hline
$\jtoss$     &86976 $\pm$ 314                  &24.1$\pm$0.7 &10.61$\pm$0.04  \\
$\ptoss$   &5447 $\pm$ 76                       &18.6$\pm$0.5 &2.52$\pm$0.04  \\
\hline
\hline
\end{tabular}
\end{center}
\label{table:br}
\end{table*}%

\section{Systematic Uncertainties}
The systematic uncertainties of the branching fraction measurements
are mainly due to the difference of efficiency between data and
simulation.  The main sources come from the difference in the
detection efficiencies of charged and neutral particles in the final
states.  In addition, the detector resolution difference between data
and MC also affects the efficiency via $\chi^{2}$ requirement in the kinematic
fit.  Other sources, such as the fitting method, parameters of
the generator and numbers of $\Psi$ events, are also considered.
Table~\ref{table:sys_br} summarizes the sources of systematic
uncertainties, which are discussed in further detail below.

\subsection{MC efficiency correction for charged tracks}
The tracking and PID efficiency differences between data and
simulation for the proton and anti-proton have been studied in bins of
transverse momentum and polar angle from control samples $\Psi \rightarrow p \bar{p} \pi^{+} \pi^{-}$.
These differences are treated as correction factors to calculate the nominal efficiencies.
The systematic uncertainties due to the limited statistics of the control samples are obtained by
summing their relative uncertainties in different bins quadratically
and are estimated to be $1.6\%$ and $1.5\%$ for $J/\psi$ and
$\psi(3686)$, respectively.

\subsection{\boldmath $\pi^{0}$ efficiency correction}
Based on control samples of $\psi(3686) \rightarrow \pi^{0} \pi^{0}
J/\psi, J/\psi \rightarrow l^{+} l^{-}$  and $e^{+} e^{-} \rightarrow \omega \pi^{0}$ at $\sqrt{s} = 3.773$ GeV, the relative difference of the $\pi^{0}$ reconstruction efficiencies between data
and MC has been obtained on the two datasets are consistent with each other.
We studied the relative difference as a function of polar angle and momentum magnitude of $\pi^{0}$, and 
found it to decrease linearly, $(0.06 - 2.41\times p)\%$, as a
function of momentum $p$.  The detection efficiency differences
obtained by varying the correction factor according to its
uncertainty, $(\sqrt{0.76\times p^{2}+1.15+0.39 \times p})\%$, are
taken as the systematic uncertainties, which are $2.3\%$ and $2.4\%$
for $J/\psi$ and $\psi(3686)$, respectively.

\subsection {Decay parameters} 
The signal MC sample is generated according to a set of decay
parameters which have been measured through multi-dimensional fitting
of angular distributions, where polarization effects and decay
asymmetry have been included~\cite{Ablikim:2020lpe}. 
Assuming no CP violation ($\alpha_{0} = \bar{\alpha}_{0}$), three
decay parameters $\alpha_{\Psi}$, $\Delta\Phi_{\Psi}$ and $\alpha_{0}$ are used for $\Psi$ decay.
We take the mean value and error matrix of these 3 parameters to build a 3 dimensional Gaussian distribution.
Based on the 3 dimensional Gaussian distribution, we generate 1000 signal MC sample to evaluate the systematic uncertainty. 
The distributions of newly obtained efficiencies are fitted
using a Gaussian function, and the widths are assigned as the
systematic uncertainties, which are 0.6\% and 0.7\% for $\jtoss$ and
$\ptoss$ respectively.

\subsection{Fitting function}
To estimate the uncertainties of the
fitting function, we use the Crystal Ball function to describe the
signal instead of the MC shape convoluted with a Gaussian function, and
the differences are taken as the systematic uncertainties, 0.4\% for
$\jtoss$ and 0.6\% for $\ptoss$.

\subsection{Background estimation}
There are two kinds of background events: peaking backgrounds and non-peaking backgrounds.
For the peaking backgrounds, we neglect this contribution in calculating the branching fractions. Considering this contributions is less than $0.1 \%$, we take $0.1\%$ as conservative estimate for this kind of systematic uncertainties.
For the non-peaking background, to estimate the uncertainties due to background modeling, we use the background shape determined by kernel density estimation from the sideband region of $M(\bar{p}\pi^{0}$) instead of the second-order polynomial function.
 The differences are taken as the systematic uncertainties, 0.9\%
 for $\jtoss$ and 0.3\% for $\ptoss$.

\subsection{$M(\bar{p}\pi^{0})$ mass window selection}
To select the signal events, we require the $1.17 < M(\bar{p}\pi^{0}) < 1.20$ $\gevcc$, which is a 30 $\mevcc$ mass window. The systematic uncertainties related to $M(\bar{p}\pi^{0})$ mass window selection are estimated by changing it to 24 $\mevcc$, 28 $\mevcc$, 32 $\mevcc$ and 34 $\mevcc$, which are corresponding to $1.172 < M(\bar{p}\pi^{0}) < 1.198$ $\gevcc$, $1.171 < M(\bar{p}\pi^{0}) < 1.199$ $\gevcc$, $1.169 < M(\bar{p}\pi^{0}) < 1.201$ $\gevcc$ and $1.168 < M(\bar{p}\pi^{0}) < 1.202$ $\gevcc$. The efficiencies and signal yields are re-evaluated with the varied windows to calculate the branching fractions. We compare the branching fractions with the nominal values and take the largest differences as the systematic uncertainties, which are 0.5\% and 0.2\% for $J/\psi$ and $\psi(3686)$ decays respectively.

\subsection{Kinematic fitting}
To estimate the systematic uncertainty caused by the $\chi^{2}_{\rm
  4C}$ requirement, we obtain the $\chi^{2}_{\rm 4C}$ distributions
using the track correction method for the helix
parameters~\cite{Ablikim:2012pg}. By imposing the requirement of
$\chi^{2}_{\rm 4C} < 100$, the efficiencies are estimated, and
compared with the nominal values. The differences, 0.1\% for both
$J/\psi$ and $\psi(3686)$, are taken as the systematic uncertainties.

\subsection{Branching fractions and numbers of $J/\psi$ and $\psi(3686)$}
The uncertainties related to the branching fractions of $\Sigma^{+}
\rightarrow p \pi^{0}$ and $\overline{\Sigma}^{-} \rightarrow
\overline{p} \pi^{0}$ are taken as $1.2 \%$ according to the
PDG~\cite{pdg}.  The numbers of $J/\psi$ and $\psi(3686)$ mesons
are determined based on inclusive hadronic events, as described in
\cite{njpsi, npsp} with an uncertainty of 0.6\% for $J/\psi$ and 0.7\%
for $\psi(3686)$.

\begin{table*}[htbp]
\caption{Systematic uncertainties }
\begin{center}
\begin{tabular}{l|c|c}
\hline
\hline
Source &$\jtoss$ (\%)   &$\ptoss$ (\%)\\
\hline
Tracking and PID efficiency   &1.6 &1.5 \\
$\pi^{0}$ reconstruction efficiency &2.3 &2.4\\
Decay parameters    &0.6 &0.7 \\
Fitting function &0.4  &0.6 \\
Peaking background &0.1 &0.1\\
Non-peaking background &0.9 &0.3 \\
$M(\bar{p}\pi^{0})$ mass window &0.5 &0.2\\
Kinematic fitting &0.1  &0.1 \\
Branching fractions &1.2 &1.2\\
Number of $\Psi$ events &0.6 &0.7\\
\hline
\hline
Total    &3.4   & 3.3\\
\hline 
\end{tabular}
\end{center}
\label{table:sys_br}
\end{table*}%

\section{Summary}
In summary, with $1310.6\times10^{6}$ $J/\psi$ and $448.1\times10^{6}$
$\psi(3686)$ events collected by the BESIII detector, the branching
fractions of $J/\psi$ and $\psi(3686)$ decaying to
$\Sigma^{+}\overline{\Sigma}^{-}$ are measured to be $(10.61 \pm 0.04
\pm 0.36) \times 10^{-4}$ and $(2.52 \pm 0.04 \pm 0.09) \times
10^{-4}$, respectively, and both are in agreement with the previous
measurement~\cite{Ablikim:2008tj, Dobbs:2017hyd} within 2 standard
deviations. The precision of the branching fraction of $\jtoss$ is
improved by a factor of 6.6 relative to the previous best measurement.  The branching fraction ratio of the
$\psi(3686)$ and $J/\psi$ decays is calculated to be $(23.8 \pm 1.1)
\%$, where the statistical and systematic uncertainties are
combined. The ratio is consistent with the previous measurement in the
$\Sigma^{0} \overline{\Sigma}^{0}$ final states by the BESIII
collaboration~\cite{previous}, and both violate the ``12\% rule''.

\acknowledgments
The BESIII collaboration thanks the staff of BEPCII and the IHEP computing center for their strong support. This work is supported in part by National Key R\&D Program of China under Contracts Nos. 2020YFA0406300, 2020YFA0406400; National Natural Science Foundation of China (NSFC) under Contracts Nos. 11625523, 11635010, 11735014, 11822506, 11835012, 11935015, 11935016, 11935018, 11961141012, 12022510, 12025502, 12035009, 12035013, 12061131003; the Chinese Academy of Sciences (CAS) Large-Scale Scientific Facility Program; Joint Large-Scale Scientific Facility Funds of the NSFC and CAS under Contracts Nos. U1732263, U1832207; CAS Key Research Program of Frontier Sciences under Contract No. QYZDJ-SSW-SLH040; 100 Talents Program of CAS; INPAC and Shanghai Key Laboratory for Particle Physics and Cosmology; Sponsored by Shanghai Pujiang Program(20PJ1401700); ERC under Contract No. 758462; European Union Horizon 2020 research and innovation programme under Contract No. Marie Sklodowska-Curie grant agreement No 894790; German Research Foundation DFG under Contracts Nos. 443159800, Collaborative Research Center CRC 1044, FOR 2359, FOR 2359, GRK 214; Istituto Nazionale di Fisica Nucleare, Italy; Ministry of Development of Turkey under Contract No. DPT2006K-120470; National Science and Technology fund; Olle Engkvist Foundation under Contract No. 200-0605; STFC (United Kingdom); The Knut and Alice Wallenberg Foundation (Sweden) under Contract No. 2016.0157; The Royal Society, UK under Contracts Nos. DH140054, DH160214; The Swedish Research Council; U. S. Department of Energy under Contracts Nos. DE-FG02-05ER41374, DE-SC-0012069

\end{document}